\documentclass[twocolumn,showpacs,superscriptaddress,prl, letterpaper]{revtex4}

\usepackage{graphicx}

\begin{document}

\title[...]
{Non-Koopmans Corrections in Density-functional Theory: Self-interaction Revisited}

\author{Ismaila Dabo}
\email{daboi@cermics.enpc.fr}
\affiliation{Universit\'e Paris-Est, CERMICS, Projet Micmac ENPC-INRIA, 
6-8 avenue Blaise Pascal, 77455 Marne-la-Vall\'ee Cedex 2, France}
\author{Matteo Cococcioni}
\affiliation{Department of Chemical Engineering and Materials Science,
University of Minnesota, Minneapolis, MN, USA}
\author{Nicola Marzari}
\affiliation{Department of Materials Science and Engineering,
Massachusetts Institute of Technology, Cambridge, MA, USA}

\begin{abstract}
In effective single-electron theories, self-interaction manifests 
itself through the unphysical dependence of the energy of an electronic state
as a function of its occupation, which results in important deviations from the ideal Koopmans trend
and strongly affects the accuracy of electronic-structure predictions. 
Here, we study the non-Koopmans behavior of local and semilocal density-functional theory (DFT) total energy methods
as a means to quantify and to correct self-interaction errors. We introduce a non-Koopmans
self-interaction correction that generalizes the Perdew-Zunger scheme,
and demonstrate its considerably improved performance in correcting the deficiencies of DFT
approximations for self-interaction problems of fundamental and practical relevance.
\end{abstract}


\maketitle

%{\it (Significance of non-Koopmans contributions in quantifying self-interaction.)}

Because density-functional theory (DFT) accounts for
correlated electron interactions via
an explicit functional $E_{\rm xc}$ of the total density $\rho$,
it provides a computational scheme that is far more predictive and practical
than the Hartree-Fock (HF) method for determining the 
Born-Oppenheimer energy of quantum systems \cite{PayneTeter1992}. 
In spite of the proven accuracy of local and semilocal DFT total energy calculations, 
it has long been recognized that the eigenvalues $\epsilon_{i\alpha}$ of the Kohn-Sham 
Hamiltonian, which are defined in the strictest sense
as the Lagrange multipliers associated to the orthonormality constraint in the self-consistent
energy minimization, have limited physical relevance \cite{ParrYang1989}. 
Indeed, due to the fact that Koopmans' theorem does not hold 
for approximate DFT functionals, the Kohn-Sham single-electron energies
are not related to any physical ionization process, at variance with their HF counterparts
that can be directly identified as the opposite electron removal energies $- \Delta E^{(0)}_{i\alpha}$ 
of the unrelaxed electronic system (i.e., without self-consistent 
modification of the electronic wavefunctions).
Following the seminal work of Perdew and Zunger 
\cite{PerdewZunger1981}, one can accurately quantify
Kohn-Sham deviations from the ideal Koopmans behavior by defining
non-Koopmans energy contributions 
$\Pi_{i\alpha} = \epsilon_{i\alpha} + \Delta E^{(0)}_{i\alpha}$ 
(this heuristic definition will be made explicit later in
the course of our discussion).

As pointed out by Perdew and Zunger, 
for any self-interaction-free single-electron theory,
the {\it Koopmans condition}
\begin{equation}
\Pi_{i\alpha}=0
\label{KoopmansCondition}
\end{equation}
is rigorously satisfied 
(in particular, the HF functional verifies $\Pi^{\rm HF}_{i\alpha}$=0 precisely because
of the cancelation of the Hartree and exchange self-energy terms) 
\cite{PerdewZunger1981}. 
This non-self-interaction criterion
establishes a central and exact correspondence between the non-Koopmans behavior of a given energy functional
and self-interaction errors, which favor electronic delocalization 
and considerably affect the accuracy of 
local and semilocal Kohn-Sham calculations.
Indeed, self-interaction is
responsible for well-known quantitative and qualitative failures of
conventional DFT functionals in describing important phenomena,
such as electron transfer \cite{SitCococcioni2006}, electronic transport 
\cite{ToherFilippetti2005}, and
electrical polarization in extended systems \cite{KorzdorferMundt2008}.

\begin{figure}[h!]
\includegraphics{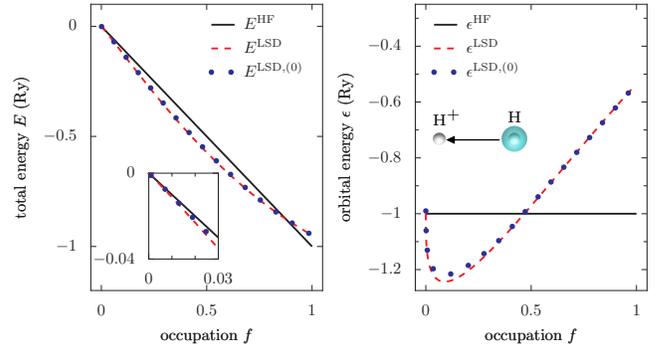}
\caption{
HF and LSD Born-Oppenheimer 
hydrogen total energies $E$ and single-electron energies $\epsilon$ with and without
electronic relaxation as the function of the fractional
occupation of the 1s orbital.
\label{HLSD}}
\end{figure}

In this work, we examine the cancelation of the non-Koopmans terms $\Pi_{i\alpha}$
that results from the Perdew-Zunger self-interaction correction (PZ-SIC)
and show that the non-self-interaction condition (Eq. \ref{KoopmansCondition})
is satisfied to second order in the single-electron densities. 
In order to achieve a higher level of accuracy,
we introduce an alternative non-Koopmans self-interaction correction (NK-SIC). 
We validate our method 
by studying the dissociation of H$_2^+$ and H$_2$ molecules, 
and compare its performance to that of the original PZ-SIC approach in describing
the longitudinal electrical response of dimerized hydrogen chains, a complex
self-interaction problem that has received much attention
for its fundamental relevance in molecular optoelectronics \cite{FaassenBoeij2002,KummelKronik2004,
BaerNeuhauser2005,UmariWilliamson2005,KorzdorferMundt2008}.

As a prelude to studying self-interaction in many-electron systems,
we analyze the non-Koopmans behavior of the local spin density (LSD) 
functional in the simplest case of the
${\rm H} \to { \rm H}^+$ ionization (Fig. \ref{HLSD}). 
The dependencies of the relaxed energy $E^{\rm LSD}$ and unrelaxed energy $E^{{\rm LSD},(0)}$ as
a function of the fractional occupation $f$ of the LSD 1s state are depicted and
compared to HF in Fig. \ref{HLSD} (in these calculations, we use the standard generalization of the HF theory 
to fractional occupations).
The derivatives of the LSD energies, which correspond to the relaxed and unrelaxed single-electron energies
$\epsilon^{\rm LSD}$ and $\epsilon^{{\rm LSD},(0)}$ (Janak's theorem), are also shown.
Since, for this one-electron system, the relaxation contribution 
$\Sigma^{\rm LSD}=E^{\rm LSD}-E^{{\rm LSD},(0)}$ vanishes
at $f=0$ and $f=1$, 
the ionization energy 
$\Delta E^{\rm LSD}=E^{\rm LSD}(0)-E^{\rm LSD}(1)$ is exactly equal to 
its unrelaxed counterpart $\Delta E^{{\rm LSD},(0)}$. 
Note however that the small relaxation energy $\Sigma^{\rm LSD}$ remains strictly negative
in the transition state region at variance with
HF, for which no relaxation occurs. Although LSD predicts 
the energy of the hydrogen atom to be $-0.94$ Ry in agreement 
with the exact HF result of $-1$ Ry, the LSD energy departs
significantly from the linear Koopmans behavior due to unphysical self-interaction:
the slopes $\epsilon^{\rm LSD}$ and $\epsilon^{{\rm LSD},(0)}$ are largely underestimated close to 
$f = 0$, reflecting the predominance of the negative $O(f^{4/3})$ exchange-correlation contribution,
while they are overestimated around $f=1$ due to the positive $O(f^2)$ Hartree term. 
Despite these notable non-Koopmans deviations, we observe that the LSD energies approach the expected 
Koopmans behavior at $f=0$ (i.e., $\epsilon^{{\rm LSD},(0)} \approx \epsilon^{{\rm LSD}} \to 
\epsilon^{\rm HF} = -1$ Ry). 
As an important consequence, the non-Koopmans
contribution to the unrelaxed ionization energy must be defined {\it by reference to
the single-electron energy of
the empty state}: $\Pi = \epsilon^{(0)}(0) + \Delta E^{(0)}$.
Extending these observations to many-electron systems with fractional occupations
and making use of Slater's theorem \cite{PerdewZunger1981}, the non-Koopmans self-interaction energies can be rewritten as
\begin{eqnarray}
\Pi_{i\alpha} & = & \epsilon_{i\alpha}^{(0)}(0)f_{i\alpha} + \Delta E_{i\alpha}^{(0)} \nonumber \\
& = & \int_0^{f_{i\alpha}} d \lambda
 \left( \epsilon_{i\alpha}^{(0)}(0) - \epsilon_{i\alpha}^{(0)}(\lambda)\right),
\label{NKEnergy}
\end{eqnarray}
where $\epsilon_{i\alpha}^{(0)}(\lambda)$ denotes the effective energy of the orbital $\psi_{i\alpha}$
with occupation $\lambda$ in the unrelaxed transition state.
Note that in the absence of self-interaction, the unrelaxed single-electron 
energies satisfy 
$d\epsilon_{i\alpha}^{(0)}(\lambda)/
d\lambda=0$ (by definition) and Eq. \ref{KoopmansCondition} is verified.

We are now in a position to evaluate the non-Koopmans energy cancelation resulting from PZ-SIC,
which consists of subtracting single-electron Hartree and 
exchange-correlation contributions to the Kohn-Sham functional $E^{\rm KS}$.
Explicitly, the Perdew-Zunger functional and orbital-dependent Hamiltonian 
are defined as
\begin{eqnarray}
E^{\rm PZ} & = & E^{\rm KS} - \textstyle\sum_{i\alpha}  E_{\rm H}[\rho_{i\alpha}] + E_{\rm xc}[\rho_{i\alpha}] 
\label{PZEnergy}
\\
\hat h_{i\alpha}^{\rm PZ} & = & 
\hat h^{\rm KS} -  v_{\rm H}[\rho_{i\alpha}] -  v_{{\rm xc}, \alpha}[\rho_{i\alpha}],
\label{PZHamiltonian}
\end{eqnarray}
where $\rho_{i\alpha}=f_{i\alpha} |\psi_{i\alpha}|^2$ denotes single-electron
densities, $v_{\rm H}=\delta E_{\rm H}/\delta \rho$ is the
electrostatic Hartree potential, and $v_{{\rm xc},\alpha}=\delta E_{\rm xc}/\delta \rho_\alpha$ 
stands for the exchange-correction potential.
Rewriting the non-Koopmans energies (Eq. \ref{NKEnergy})
in terms of the corrected Hamiltonian (Eq. \ref{PZHamiltonian}) and
expanding each energy contribution in terms of the variables $\rho_{i\alpha}$
at the self-interaction-free empty-state
density $\overline{\rho_{i\alpha}}=(\rho_\alpha-\rho_{i\alpha},\rho_{\overline \alpha})$ 
(with $\overline{ \alpha}$ opposite to $\alpha$), we obtain:
\begin{eqnarray}
\Pi^{\rm PZ}_{i\alpha} & = & \displaystyle\int_0^{f_{i\alpha}} d\lambda
\langle \psi_{i\alpha}  |
\left( \hat h_{i\alpha}^{\rm PZ}(0)
- \hat h_{i\alpha}^{\rm PZ}(\lambda) \right) | \psi_{i\alpha} \rangle \nonumber \\
& \propto & \sum_{j\neq i} \displaystyle\int
f_{{\rm xc},\alpha\alpha\alpha}^{(3)}[\overline{\rho_{i\alpha}}]\rho_{j\alpha}
\rho_{i\alpha} \rho_{i\alpha}  d{\bf r}_1 d{\bf r}_2 d{\bf r}_3,
\label{PZCancellation}
\end{eqnarray}
where $f^{(n)}_{\rm xc}[\rho]$ denotes the $n$th-order 
functional derivative of $E_{\rm xc}[\rho]$.
As a result, PZ-SIC cancels
self-interaction contributions
up to the second order in the single-electron densities, which 
considerably ameliorates the precision of calculated ionization energies
and related properties \cite{PerdewZunger1981,KummelKronik2008} but
tends to sensibly overestimate 
self-interaction errors for many-electron systems \cite{FilippettiSpaldin2003,
VydrovScuseria2006,BylaskaTsemekhman2006}.

A more accurate self-interaction cancelation
can be obtained by modifying the expression 
of the corrective terms in Eq. \ref{PZEnergy}. 
In fact, due to the direct correspondence between 
non-Koopmans contributions and self-interaction errors, it is legitimate
to identify the orbital self-interaction correction
as the single-electron non-Koopmans energy $\Pi^{\rm KS}_{i\alpha}$. 
The resulting NK-SIC energy reads
$E^{\rm NK} = E^{\rm KS} + \sum_{i\alpha} \Pi^{\rm KS}_{i\alpha}$
and variation with respect to the single-electron density $\rho_{i\alpha}$ yields
the orbital-dependent Hamiltonian
\begin{eqnarray}
\hat h_{i\alpha}^{\rm NK} & =  & \hat h^{\rm KS} - v_{\rm H}[\rho_{i\alpha}]
+ v_{{\rm xc},\alpha}[\overline{\rho_{i\alpha}}]
- v_{{\rm xc},\alpha}[\rho]
+ w_{{\rm xc},i\alpha} \nonumber \\
& = & -\textstyle\frac{1}{2} \nabla^2 + v + v_{\rm H}[\overline{\rho_{i\alpha}}] 
+ v_{{\rm xc},\alpha}[\overline{\rho_{i\alpha}}] + w_{{\rm xc},i\alpha},
\label{NKHamiltonian1}
\end{eqnarray}
where $v$ is the external potential and $w_{{\rm xc},i\alpha}$ is defined as
\begin{eqnarray}
w_{{\rm xc},i\alpha}({\bf r}) & = & \textstyle\sum_{j\beta\neq i\alpha} v_{{\rm xc},\alpha}([\overline{ \rho_{j\beta}}];{\bf r})
- v_{{\rm xc},\alpha}([\rho];{\bf r}) \nonumber \\
& & + \int d{\bf r}' f^{(2)}_{{\rm xc},\alpha \beta}([\overline{\rho_{j\beta}}];{\bf r},{\bf r'}) \rho_{j\beta}({\bf r}').
\label{NKHamiltonian2}
\end{eqnarray}
Substituting Eqs. \ref{NKHamiltonian1} and \ref{NKHamiltonian2} in the expression
of the non-Koopmans energies (Eq. \ref{NKEnergy}), we obtain
the following estimate for the self-interaction contributions:
\begin{eqnarray}
\Pi_{i\alpha}^{\rm NK} \propto \sum_{j\beta\neq i\alpha}  \displaystyle\int
 f_{{\rm xc},\alpha\alpha\beta\beta}^{(4)}([\overline{\rho_{j\beta}}];{\bf r}_1,{\bf r}_2,{\bf r}_3,{\bf r}_4 )
\nonumber \\
\times
\rho_{j\beta}({\bf r}_4)\rho_{j\beta}({\bf r}_3)
\rho_{i\alpha}({\bf r}_2) \rho_{i\alpha}({\bf r}_1) 
 d{\bf r}_1  d{\bf r}_2  d{\bf r}_3 d{\bf r}_4
\end{eqnarray}
(note that only the term $w_{{\rm xc},i\alpha}$ contributes to
self-interaction).
Consequently, our modification of the PZ-SIC functional brings about
an additional factor of precision in the cancelation of spurious electronic interactions.
The improved accuracy of the NK-SIC scheme in comparison to PZ-SIC will be shown below.

At this point, one may observe that an exact cancelation of the
non-Koopmans self-interaction contributions is obtained by defining the orbital-dependent
Hamiltonian as $\hat h_{i\alpha} = 
-\frac{1}{2} \nabla^2 + v + v_{\rm H}[\overline{\rho_{i\alpha}}] 
+ v_{{\rm xc},\alpha}[\overline{\rho_{i\alpha}}] = -\frac{1}{2} \nabla^2 + v_{i\alpha}[\overline{\rho_{i\alpha}}]$. 
Nevertheless, such a scheme does not derive from the minimization of any physical
energy, as this would lead to the following variational inconsistency:
$\frac{\partial^2 E}{ \partial \rho_{i\alpha}
\partial \rho_{j\beta}} = \frac{\partial v_{j\beta}}{\partial \rho_{i\alpha}} \neq 
\frac{\partial v_{i\alpha}}{\partial \rho_{j\beta}} = \frac{\partial^2 E }{ \partial \rho_{j \beta}
\partial \rho_{i \alpha}}$.
The NK-SIC Hamiltonian (Eq. \ref{NKHamiltonian1}), which
includes the additional term $w_{{\rm xc},i\alpha} =O(\sum_{j\beta\neq i\alpha} \| \rho_{j\beta}^2 \|)$
for variationality, 
provides a close {\it physical} approximation to this exact correction.

The implementation of the NK-SIC method requires computing the kernel of the
local or semilocal exchange-correlation energy $f_{\rm xc}^{(2)}$.
This calculation is performed using the scheme proposed by 
Dal Corso and de Gironcoli in the context of phonon-dispersion
computations \cite{DalCorsodeGironcoli2000}. Moreover, the determination of the NK-SIC
terms necessitates calculating the Hartree self-interaction energy $E_{\rm H}[\rho_{i\alpha}]$ and potential
$v_{\rm H}[\rho_{i\alpha}]$,
which entails treating a system with a net electrical charge. 
To eliminate periodic-image errors in the plane-wave evaluation of $E_{\rm H}[\rho_{i\alpha}]$ 
and $v_{\rm H}[\rho_{i\alpha}]$, we employ
countercharge correction methods \cite{DaboKozinsky2008}.
Since the determination of
$f^{(2)}_{\rm xc}$ and the periodic-image corrections are relatively inexpensive, the computational burden
of NK-SIC computations is comparable to that
of conventional PZ-SIC calculations.

\begin{figure}
\includegraphics{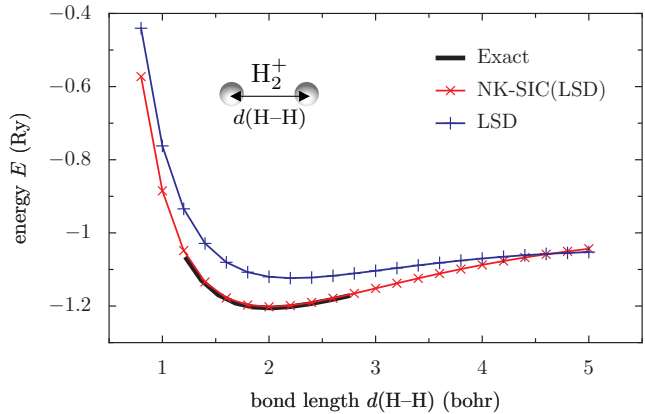}
\caption{
Exact, NK-SIC(LSD), and LSD potential energy curves of ${\rm H}_2^+$ as
a function of the H--H bond length.
\label{H2+}}
\end{figure}

We first study the dissociation of the H$_2^+$ molecular ion.
The LSD and NK-SIC(LSD) energies are compared to
the exact Born-Oppenheimer potential energy curve in Fig. \ref{H2+}.
We observe that the LSD energies of this one-electron system are largely overestimated by as much as 0.1 Ry
in the vicinity of the equilibrium bond length due to self-interaction errors, at variance with
the NK-SIC energies, which are in close concordance with the exact results. Moreover, in the 
infinite separation limit, the NK-SIC electronic ground state
is correctly predicted to be degenerate (with an energy of $-1$ Ry)
with respect to the effective occupations $n_1$ and $n_2=1-n_1$ of the 
two hydrogen sites, 
at variance with LSD, which yields
a nondegenerate and delocalized split-electron state ($n_1=n_2=1/2$) of
lower energy. 

\begin{figure}
\includegraphics{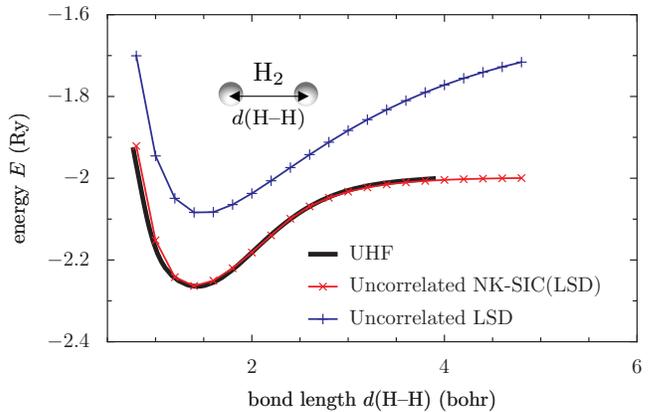}
\caption{
UHF, NK-SIC(LSD), and LSD potential energy curves of ${\rm H}_2$
in the absence of electronic correlation as a function of the H--H bond length.
\label{H2}}
\end{figure}

As a second self-interaction paradigm, we examine the
dissociation of H$_2$ in the uncorrelated limit.
The dependencies of the unrestricted Hartree-Fock (UHF), uncorrelated NK-SIC(LSD), and 
uncorrelated LSD Born-Oppenheimer energies as a function of the intramolecular distance 
are depicted in Fig. \ref{H2}.
Similarly to the H$_2^+$ case, we observe
that LSD overestimates the total energy of the system,
in contrast to NK-SIC, which brings the potential energy curve
in close accordance with the self-interaction-free result. 
Furthermore, the analysis of the spin occupations $n_{i\alpha}$ of the hydrogen sites
at large atomic separation reveals that both the NK-SIC and LSD methods predict a two-fold degenerate 
ground state, in which each electron is localized at one
site ($n_{1\uparrow}=n_{2\downarrow}=1$ or $n_{1\downarrow}=n_{2\uparrow}=1$). 
However, this transition occurs for distinct reasons 
in the two cases: while electronic localization
is driven by the reduction of the interelectronic Hartree energy for NK-SIC,
it is the enhancement of unphysical self-exchange that leads to localization
for LSD. This fact explains salient differences in the asymptotic behaviors of
the LSD and NK-SIC energies.

\begin{table}
\caption{
Longitudinal polarizability $\alpha_N$ (in a.u.) of the hydrogen chain as a function of the number of
dimers $N$.
The mean absolute deviations $\Delta \alpha$ from CCSD(T) results are also reported. 
\label{HChain}}
\begin{tabular}{lrrrrr}
\\
\hline \hline
method ~~~~~~~ & ~~~~~~~ $\alpha_2$ & ~~~~~~~ $\alpha_3$ & ~~~~~~ $\alpha_4$ & 
~~~~~~ $\alpha_6$ &
~~~~ $\Delta \alpha$ \\
\hline 
CCSD(T)\footnotemark[1]& 29.0 & 50.9 & 74.4 & 124.0 & ---  \\
MP4\footnotemark[2] & 29.5 & 51.6 & 75.9 & 126.9 & 1.4 \\
NK-SIC(LSD) & 30.9 & 49.0 & 71.3 & 122.3 & 2.1 \\
HF\footnotemark[1]& 32.2 & 56.6 & 83.0 & 138.6 & 8.0 \\
PZ-SIC(LSD)\footnotemark[1] & 33.0 & 59.7 & 89.1 & 152.0 & 13.9 \\
LSD\footnotemark[3] & 37.6 & 72.9 & 115.3 & 212.1 & 39.9 \\
\hline \hline
\end{tabular}
\flushleft
\footnotemark[1]{Ref. \cite{RuzsinszkyPerdew2008}},
\footnotemark[2]{Ref. \cite{ChampagneMosley1995}},
\footnotemark[3]{Ref. \cite{KummelKronik2004}}.
\end{table}

Having validated the NK-SIC method for archetypical one- and two-electron problems (for 
which PZ-SIC would also be exact), 
we turn to the difficult case of the electrical response
of extended hydrogen chains, which are frequently employed as model systems
in molecular optoelectronics and represent a critical test
in assessing the predictive performance of electronic-structure methods \cite{RuzsinszkyPerdew2008}.
We consider a linear chain geometry
characterized by alternating H--H distances of 2 and 3 bohr, 
as reported in the literature \footnote{We employ a norm-conserving hydrogen pseudopotential with an
energy cutoff of 25 Ry for the plane-wave expansion of the electronic wavefunctions.
The NK-SIC problem is solved using a preconditioned conjugate-gradient
band-by-band method \cite{PayneTeter1992}.
The longitudinal electrical 
polarizability $\alpha$ is evaluated by finite difference
from the dipole moment computed at 
an electric field of 0.005 a.u., starting from zero electric field.
Using the above method and calculation parameters,
the polarizability of an isolated hydrogen atom is verified to be 
within less than half a percent of the exact value ($\alpha({\rm H})=4.5$ a.u.).}.
The LSD and NK-SIC longitudinal polarizabilities $\alpha_N$ of the hydrogen chain
as a function of the number of H$_2$ units are compared with HF, fourth-order M{\o}ller-Plesset perturbation
theory (MP4), coupled cluster [CCSD(T)], and PZ-SIC predictions
in Table \ref{HChain}. We observe that the computed polarizabilities follow 
an increasing and progressively linear trend as a function of the number of hydrogen dimers.
In spite of these similarities, 
we note that LSD dramatically overestimates the electrical response of the system.
The Perdew-Zunger correction sensibly improves the accuracy of LSD results,
reducing the mean absolute error from 
$\Delta \alpha^{\rm LSD}=39.9$ to $\Delta \alpha^{\rm PZ}=13.9$ a.u.
relative to CCSD(T) polarizabilities (which we choose as our standard of accuracy).
Despite this notable amelioration, the performance of the PZ-SIC scheme remains inferior to that of 
self-interaction-free HF ($\Delta \alpha^{\rm HF}=8.0$ a.u.). This should be contrasted to
NK-SIC predictions, which are found to be much more accurate than their HF counterparts with a precision
comparable to the mean absolute difference between post-Hartree-Fock
CCSD(T) and MP4 calculations ($\Delta \alpha^{\rm NK}=2.1$ a.u.).
These findings confirm the self-interaction origin of the overestimated
polarizabilities within LSD and 
provide a clear illustration of the superior predictive ability of the NK-SIC method
for complex self-interaction problems.

As a final note, it is important to mention that several successful refinements of the PZ-SIC method have been
proposed in the literature \cite{VydrovScuseria2006,BylaskaTsemekhman2006}. 
One limitation of such approaches is that they require
to scale down the Hartree self-energy although this corrective term is in principle exact (as
a result of the linearity of $v_{\rm H}[\rho]$). 
The NK-SIC method represents a parameter-free alternative to 
semi-empirical downscaling procedures.

In conclusion, we have revisited self-interaction in light of the
Koopmans condition (Eq. \ref{KoopmansCondition}) and
introduced a non-Koopmans correction (NK-SIC), which 
provides  a more accurate cancelation of self-interaction contributions in comparison
to PZ-SIC.
In order to validate the NK-SIC method, we have studied the
dissociation of H$_2^+$ and H$_2$, finding excellent agreement with exact and self-interaction-free
calculations. Finally, we have demonstrated the much
improved performance of the NK-SIC scheme
in describing the electrical response of hydrogen chains, which represents a
stringent and technologically relevant benchmark for electronic-structure methods. 

The computations in this work have been performed using the 
Quantum-Espresso package \cite{Espresso}. 
The authors acknowledge support from the Grant in Aid for Research, Artistry and Scholarship of the 
Graduate School of the University of Minnesota. 
Helpful suggestions and comments from Stefano de Gironcoli, \'Eric Canc\`es, 
Nicolas Poilvert, and Andrea Floris are gratefully acknowledged.

\bibliography{article}

\end{document}